\documentclass{aip-cp}

\usepackage[numbers]{natbib}
\usepackage{rotating}
\usepackage{graphicx}
\usepackage{lineno}

\begin{document}

\def\nuc#1#2{${}^{#1}$#2}
\newcommand{\nonubb}  {$0 \nu \beta \beta$}

\title{A Review and Outlook for the Removal of Radon-Generated Po-210 Surface Contamination}

\author[usc]{V.E.~Guiseppe\corref{cor1}}
\author[sdsmt]{C.D. Christofferson}
\author[usc]{K.R. Hair}
\author[usc]{F.M. Adams}
\affil[usc]{Department of Physics and Astronomy, University of South Carolina, Columbia, SC, USA}
\affil[sdsmt]{South Dakota School of Mines and Technology, Rapid City, SD, USA}
\corresp[cor1]{Corresponding author: guiseppe@sc.edu}

\maketitle


\begin{abstract}
The next generation low-background detectors operating deep underground aim for unprecedented low levels of radioactive backgrounds. 
The deposition and presence of radon progeny on detector surfaces is an added source of energetic background events. 
In addition to limiting the detector material's radon exposure in order to reduce potential surface backgrounds, it is just as important to clean surfaces to remove inevitable contamination.  Such studies of radon progeny removal have generally found that a form of etching is effective at removing some of the progeny (Bi and Pb), however more aggressive techniques, including electropolishing, have been shown to effectively remove the Po atoms. In the absence of an aggressive etch, a significant fraction of the Po atoms are believed to either remain behind within the surface or redeposit from the etching solution back onto the surface. We explore the chemical nature of the aqueous Po ions and the effect of the oxidation state of Po to maximize the Po ions remaining in the etching solution of contaminated Cu surfaces. We present a review of the previous studies of surface radon progeny removal and our findings on the role of oxidizing agents and a cell potential  in the preparation of a clean etching technique. 
\end{abstract}

\section{INTRODUCTION AND MOTIVATION}
Observation of rare processes demands a careful study and implementation of signal processing and detector design to reach new levels of background suppression. Both the potential backgrounds and detector response to backgrounds must be thoroughly understood. Radon (Rn) gas provides a problematic class of backgrounds due to its terrestrial presence and its long-lived progeny. Exposure to radon at any stage of assembly of an experiment can result in surface contamination by progeny supported by the long half-life (22 yr) of \nuc{210}{Pb} on sensitive surfaces of a detector \cite{ams07, leu05,gui11}. The radon progeny surface contaminants can continue to produce unwanted background even after the detector is moved to its final laboratory or configuration with a lower radon level. In the case of neutrinoless double-beta decay experiments, energetic  $\alpha$ and $\beta$ decays of \nuc{210}{Po} and \nuc{210}{Bi}, respectively, produce a background near or above the region of interest. Particularly problematic is the degraded alpha deposit that can fall under the region of interest requiring pulse shape or fiducial cuts to be rejected. In the case of a dark matter detector, the $\alpha$ decay of \nuc{210}{Po} produces a nuclear recoil similar to that of a WIMP-nucleon scatter. Because of the importance of this class of backgrounds, all low background rare event searches pay close attention to Rn exposure and are exploring ways to mitigate and control Rn-generated backgrounds. 

While the Rn progeny deposit directly on a material surface, the nuclear recoils of the alpha decay of \nuc{218}{Po} and \nuc{214}{Po} cause the daughters to be implanted deeper into the surface. Modeling and surface alpha measurements 
indicate that the Rn-borne \nuc{210}{Pb} is implanted down to 0.05 -- 0.1 $\mu$m in metals due to the deposition of progeny earlier in the chain. Further, exposure to Rn gas invites diffusion of Rn that can carry progeny deeper into the surface layers of material. The diffused  Rn progeny  and any possible bulk contamination of \nuc{210}{Pb} creates an additional path of material for the degradation of an emitted alpha background. Though the majority of alpha emission may come from deposited daughters close to the surface, the surface roughness of a material adds an additional layer of material an alpha must traverse \cite{per13}. 
The net effect is an emitted alpha with a peak near its full energy, but with a low energy tail extended down to low energies as if the alpha originated deeper than the implantation depth of 0.05 - 0.1 $\mu$m. For a typical metal, the range of a 5 MeV alpha is 20 $\mu$m. One method to mitigate surface alpha contamination is to chemically remove the Rn progeny that exist down to a surface depth that they are imbedded. 

The deposition of Rn progeny continues to be thoroughly modeled and studied by low background experiments. The findings are helping establish the infrastructure and facility designs needed for future experiments to control the exposure to Rn gas. Naturally, the deposition studies are complimented by separate studies evaluating the cleaning and surface removal of Rn progeny.  Since the start of the Low Radioactivity Techniques workshop series, there have been over 10 papers focusing on the modeling, deposition, or removal of surface Rn progeny \cite{leu05,woj07,gui11,woj11,jil13,pat13,per13,sch13,bru15,kob15,zuz15}, not to mention the papers where Rn studies are described in a more general experimental program overview paper or studies reported elsewhere. The findings of surface progeny removal studies have shown mixed results. Generally, the Pb and Bi progeny has been found to be easily removed using a variety of standard cleaning methods. However, in some materials, Po has been more difficult to remove and more aggressive chemical cleaning methods are often recommended. 

The mixed results for the removal of surface \nuc{210}{Po} are worrisome given the needs of next generation low background, rare event searches. With larger experiments, there will be a greater number of parts with a more stringent surface contamination control required. When deciding on a cleaning procedure, next generation experiments must consider: 1) efficiencies of removing Rn progeny, 2) quantities and purities of chemicals needed, 3) chemical waste, 4) generation and mitigation of chemical fumes, 5) underground cleaning limitations and logistics, 6) number of parts and process automation compatibility, and 6) maintaining dimensional and mechanical tolerances.

\section{Po REMOVAL TECHNIQUES}

Surface cleaning techniques have been evaluated with varying results. In this section, we describe some of the methods used to highlight the generally accepted findings. The work of \citet{hop07} looked at the cleaning of Cu and evaluated two methods. The Cu samples used in the study were initially loaded with $^{209}$Po by electrodeposition and the removal of the Po atoms was used to evaluate the various surface cleaning methods. The most effective treatment of removing Po surface contamination was a concentrated nitric acid etch. Lowering the concentration of nitric also lowered the effectiveness of Po removal. In the paper, the authors also introduce an alternate acidified peroxide cleaning method containing a dilute piranha solution of 1\% H$_2$SO$_4$ and 3\% H$_2$O$_2$ (hereinafter referred to the PNNL solution for cleaning Cu). This PNNL solution can effectively etch away surface layers in a more controlled fashion than concentrated nitric solution. Though it is shown to be effective at removing Cu atoms from the surface, it has limited effect of removing Po surface contamination. The authors stated some concern with Po solubility in the solution given they found improved Po removal when starting with Cu samples of lower initial Po activity. 

Similar tests were conducted by other groups and two studies are worth mentioning here. \citet{zuz15} looked at the removal of all three long-lived Rn progeny ($^{210}$Pb, $^{210}$Bi, $^{210}$Po) from samples of Cu, stainless steel and Ge. Using the standard PNNL acidified peroxide solution, the authors found that cleaning Cu samples showed better reduction of Bi and Pb than of Po, which showed very little, if any, reduction. Explanations for the poor removal of Po, including redeposition in solution, are given in Ref. \cite{zuz12}. The removal of Rn progeny from contaminated samples of steel and Ge faired a bit better than Cu indicating the substrate material has a significant effect on contamination removal efficiency. The authors additionally tried a separate electropolishing technique on  Cu and stainless steel. The electropolishing treatment significantly improved the removal efficiency of all three progeny, including a significant improvement for Po. The next study given in Ref.  \cite{sch13} also found effective reduction of Po through an electropolishing procedure on stainless steel. Interestingly, this study reduced Po to detection sensitivity levels after a depth of 0.6 $\mu$m steel was removed, while the implantation depth of Rn progeny is modeled to be 0.05 - 0.1 $\mu$m in most metals. 

The fact that methods to remove Po have had mixed results indicates the importance of understanding the chemical behavior of Po in solution and the substrate being cleaned. 
Most methods tried using some form of an acidic solution, but given the behavior of Po in the Pourbaix diagram (see Ref. \cite{ans12}), neutral Po can exist in solution over the entire pH range. In the neutral state, Po will likely redeposit onto a surface. However at low pH, Po can be driven into the Po(IV) oxidized state. When in a  ion state, Po should favor staying in solution rather than redepositing based on the aqueous solution chemistry. The Po can be converted into a stable ion state by applying an oxidation potential or an oxidizing agent.  The standard electrode potentials for Po in an acidic solution suggests that a potential of 0.73 V separates neutral Po from Po$^{4+}$, which is expected to be the most stable Po ion in solution \cite{ans12}.

Working off the scenario that the ion state of Po is a significant factor in its removal from a contaminated sample, the following requirements are needed. First, the Po must be converted into an ion state through oxidation and favorably into the Po$^{4+}$ state. Solubility  plays a role so a sufficient amount of surface area should be given within a cleaning solution for the amount of Po present. Finally, the stability of the ion state must be maintained to avoid reduction back to a neutral atom. The substrate atoms  (e.g Cu) will likely play a role as they compete for the oxidizer (either an applied potential or an oxidizing agent). Given the separate oxidation potentials, the kinetics of the Po and the substrate atom removal may occur at different rates. That is, the depth of the substrate that is removed may not necessarily mean all other species located within that depth have had time to be oxidized before redepositing back onto the surface. Instead, sufficient exposure to the oxidizer may be the determining factor of effective Po removal, rather than the depth of the substrate removed. Greater exposure can be accomplished several ways: the contaminated sample can be agitated to make use of the volume of solution exposure, greater concentrations of an oxidizing agent, longer exposure times, larger volumes of solution surrounding the contaminated sample, etc. 

Returning to the mixed results for Po noted earlier, we can now put the results in the context of the Po ion state scenario. The tests with nitric acid \cite{hop07} worked very well for all progeny. Nitric acid is an excellent oxidizer and is aggressive to almost all metals. A problem with concentrated nitric is that it will remove a large amount of material in the process in an uncontrolled way due to thermodynamic changes during dissolution.  Electropolishing was found to be very efficient in removing Po and the other progeny \cite{zuz15,sch13}. Electropolishing recipes typically call for concentrated acids to keep the pH low and an applied electric potential provides the necessary oxidation. This method drives off a large portion of the surface layers, which could compromise mechanical properties if used on small parts. The PNNL acidified peroxide solution uses H$_2$0$_2$ as the oxidizing agent. With this method, there is a modest and controlled material removal, which is an important consideration for tight dimensional tolerances on small parts. Though this solution meets some of the the requirements for removal of Po, the previous tests had found this method does not effectively remove Po from  Cu. It is possible the method can be effective with increased exposure to the oxidizing agent, which is what we explore and test in the next section. 

\section{Po REMOVAL STUDY}

For our study, we wish to explore the hypothesis of the role played by oxidation in removing Po from a sample. Given previous studies have shown the poor removal of Po from Cu, we will focus only on quantifying the removal of Po from Cu samples. Cu foil samples 50 mm in diameter and 0.5-mm thick  are exposed to a 100 kBq radon source for about 1 month. The samples are left for over 2 years to allow the $^{210}$Po activity to grow-in to equilibrium with the $^{210}$Pb. The net surface alpha rate achieved is around 300 cts/day (not corrected for detection efficiency). An alpha spectrometer using an ion-implanted silicon detector is used to count the Cu samples before and after chemical treatment. The alpha detector has a background count rate of 6 counts/day. The goals of this study are to explore certain features of chemical treatment to assess the role played by oxidizers on the removal of Po from the contaminated surface. Based on the ratio of the pre-treatment and post-treatment alpha rates, the fraction of Po removed is determined from the background-subtracted alpha rates, which is not dependent  on the detector counting efficiency since the geometry remains unchanged. 

\section{RESULTS AND DISCUSSION}

\begin{figure}[t]
\centering
\begin{tabular}[b]{c}
  \includegraphics[width=0.49\textwidth]{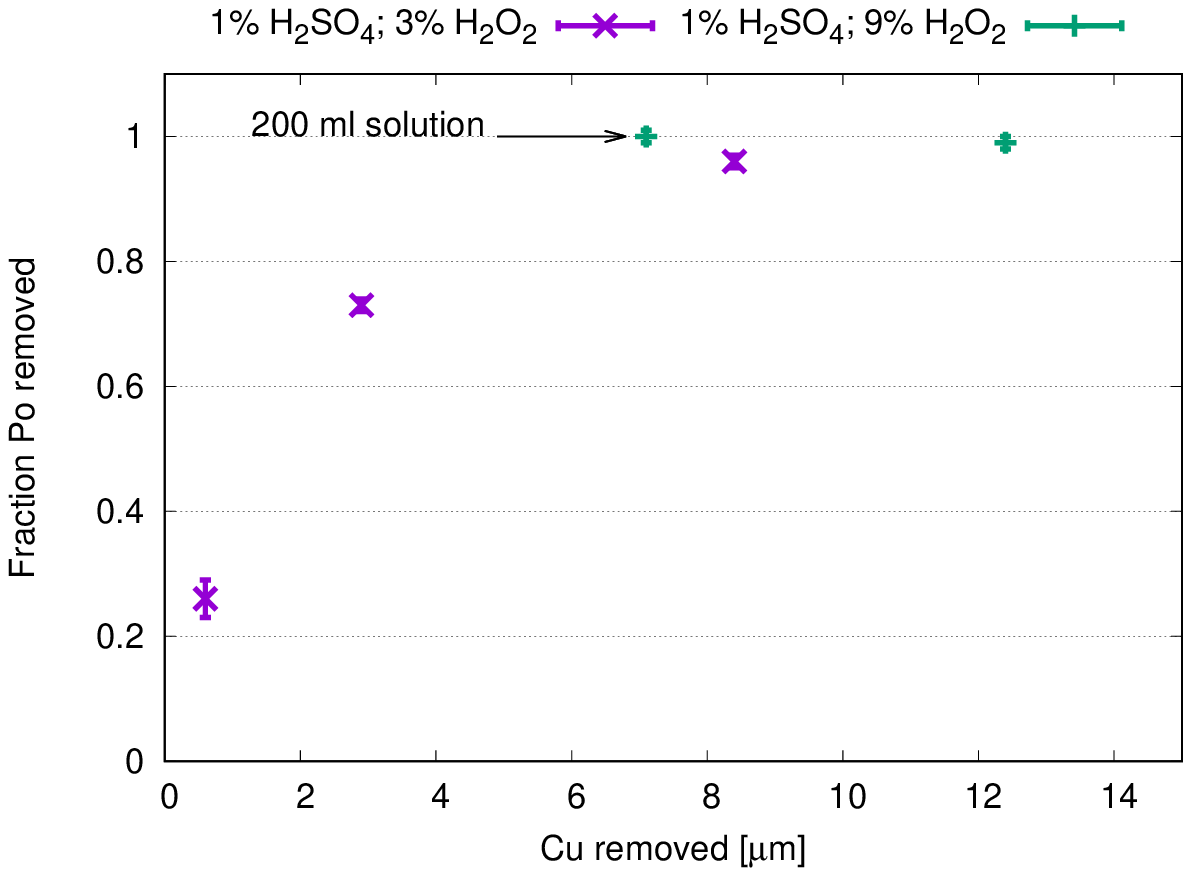} \\
  \small (a)
\end{tabular}
 \quad
\begin{tabular}[b]{c}
  \includegraphics[width=0.49\textwidth]{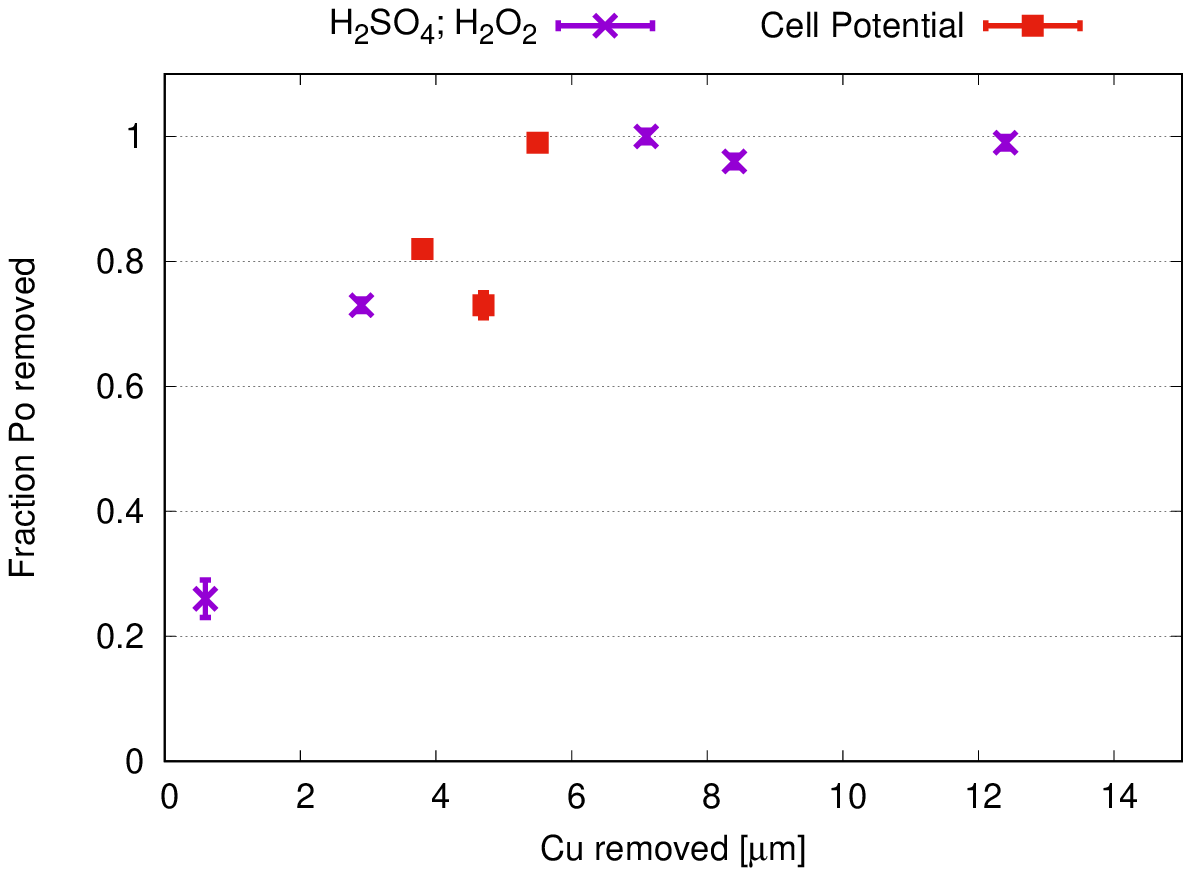}\\
  \small (b) 
  \end{tabular}
  \caption{(a) The fraction of Po removed from a cleaning solutions plotted against the depth of Cu removed. The standard PNNL acidified peroxide solution is used for a varying amount of time, though the H$_2$0$_2$ is increased for one set of samples. (b) The same data points from (a), but with an additional set where a cell potential is applied.}
  \label{fig:1}
\end{figure}

The first test was to simply look at the effect of increasing the H$_2$0$_2$ concentration from 3\% to 9\% but otherwise following the PNNL acidified peroxide solution process. Each sample was exposed to a fresh solution for a varying amount of time to achieve a range of Cu thickness removed from each face of the sample. The samples were agitated equally during all tests and all but one test was performed with a 100 ml solution. The results in Fig. \ref{fig:1}(a) show a general trend of improved Po removal with greater Cu removal up to the maximum 100\% Po removal (within counting uncertainties consistent with the background rate of the alpha detector).  However, one should not generally conclude that the depth of Cu removed down to the expected Po implantation depth should determine when all of the Po is removed. Recall, the implanted Po should exist down to 0.05 $\mu$m.  Instead, the amount of Cu removed is likely a proxy for quantifying the exposure to the oxidizing agent converting Po into the Po$^{4+}$ ion state.  The conditions present may allow the Cu removal to indicate when sufficient exposure for removal of the Po has occurred. Further, the greater concentration of H$_2$0$_2$ does not have a strong effect on the Po removal; but it does give sufficient exposure in a shorter time as confirmed by the amount of Cu removed. 

To test if a cell potential could aid in the oxidation of Po in the same acidified peroxide solution, three additional samples were cleaned with an applied potential greater than 0.73 V. The results of the three tests are given in Fig. \ref{fig:1}(b) and plotted with the same non-potential tests from Fig. \ref{fig:1}(a), where there is no longer a distinction made for the concentration of H$_2$0$_2$. The results show no added benefit of adding a cell potential, which suggests the oxidizing agent present is the dominant oxidizing mechanism under these conditions. Instead, the additional samples follow the same general trend of increasing Po removal with greater exposure, as indicated by the increased Cu removal. With the additional samples present, the variability among samples is more visible when plotted against exposure. This observation indicates another variable may be controlling the effectiveness of Po removal.

\begin{figure}[ht]
\centering
\begin{tabular}[b]{c}
  \includegraphics[width=0.49\textwidth]{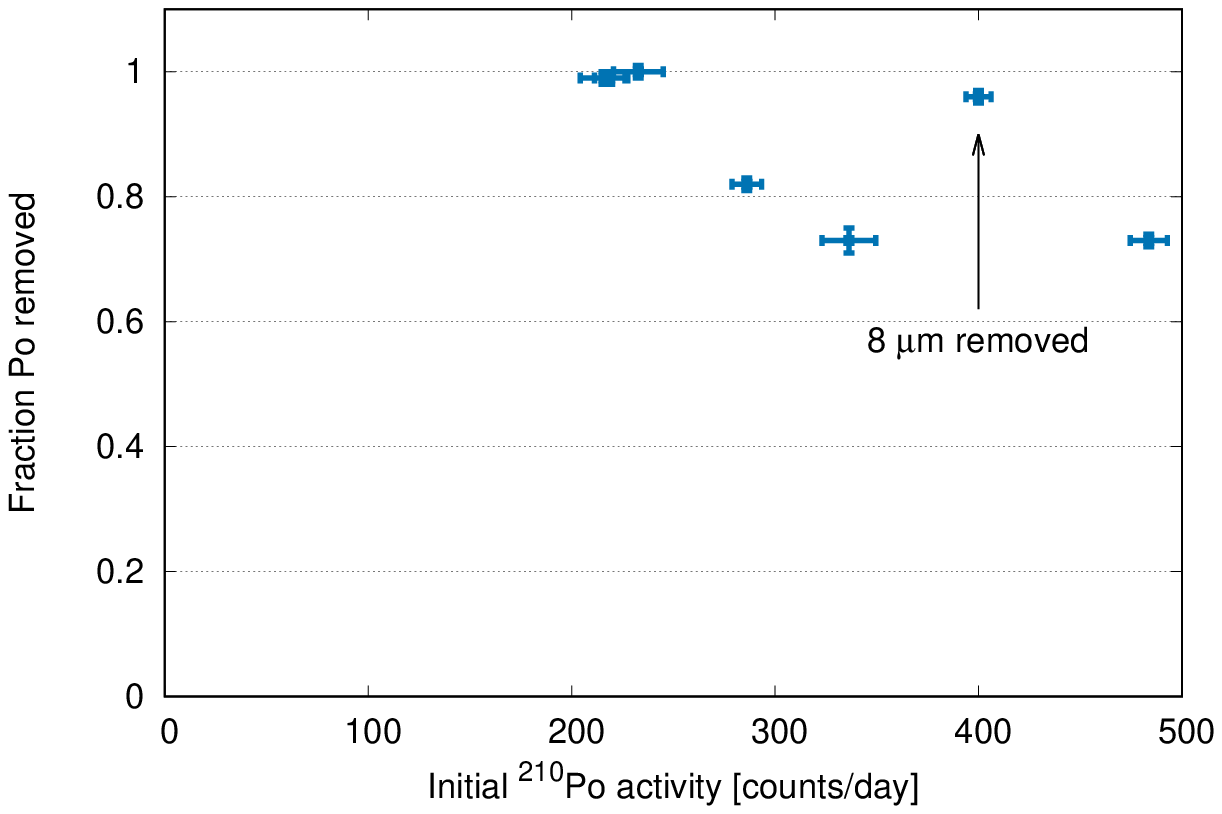} \\
  \small (a)
\end{tabular}
 \quad
\begin{tabular}[b]{c}
  \includegraphics[width=0.49\textwidth]{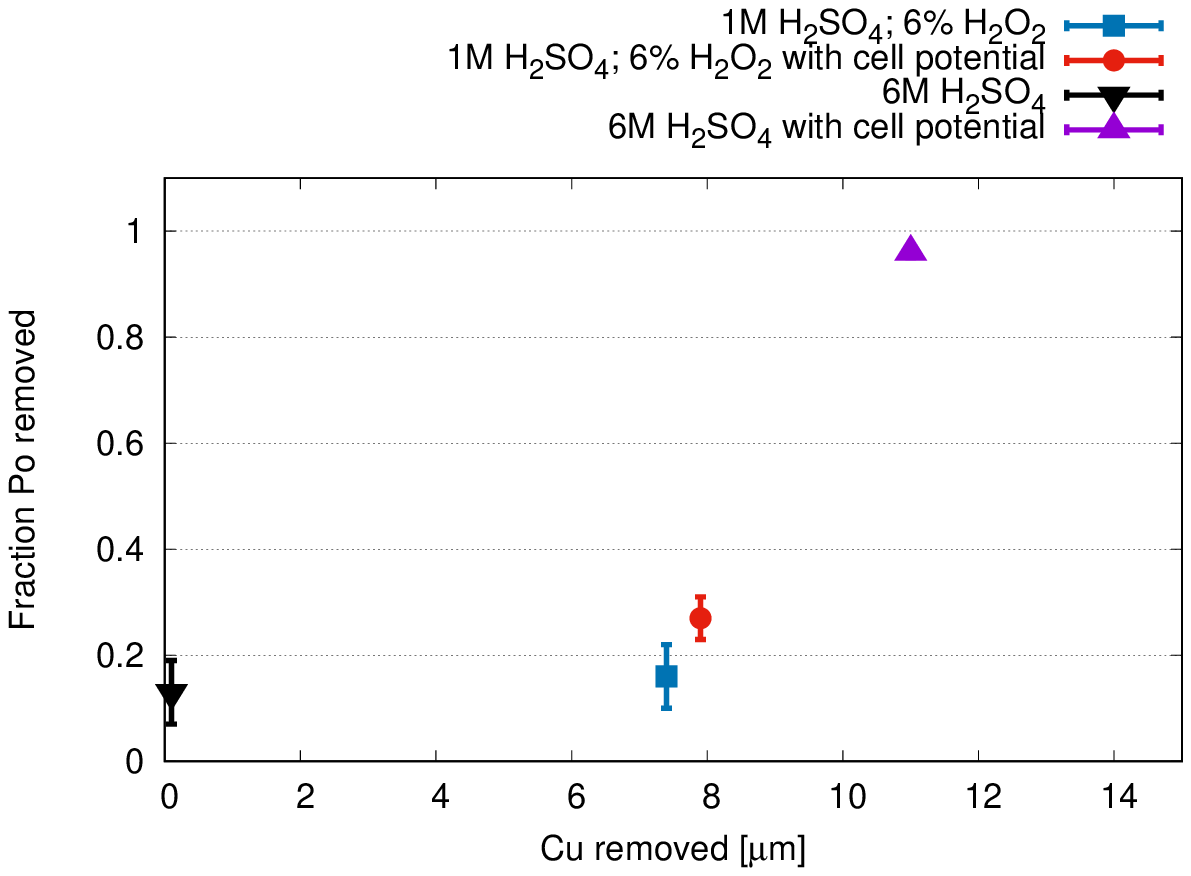}\\
  \small (b) 
  \end{tabular}
  \caption{(a) The fraction of Po removed from a cleaning solutions plotted against the initial $^{210}$Po surface activity. for the samples in Fig. \ref{fig:1} where $>2 \mu$m of Cu was removed. See the discussion in the text regarding the sample with an activity of 400 cts/day. 
  (b) The fraction of Po removed from a cleaning solutions plotted against the depth of Cu removed for two alternate cleaning methods. In one set, an acidified peroxide solution of lower pH  is used with and without an applied cell potential resulting a small fraction of Po removal. In the second set, only a stronger acidic solution is used with and without an applied cell potential. Removal of Cu and Po is found with the application of a cell potential.
  }
  \label{fig:2}
\end{figure}

The data from Fig. \ref{fig:1}(b) with the most effective Po removal (where $>2\, \mu$m of Cu removed) is plotted in Fig. \ref{fig:2}(a) as a function of initial $^{210}$Po surface activity.  The fraction of Po removed appears to generally  increase as the initial $^{210}$Po activity decreases, with one notable exception. The sample with a starting activity of 400 cts/day had an acidified peroxide exposure sufficient to remove 8 $\mu$m of Cu while the other two samples with starting activity $>$300 cts/day had exposures removing $<6\,\mu$m of Cu. The results suggest the initial Po activity dictates the exposure to the solution required. When the exposure to the oxidizing agent is increased, commensurate with increased initial $^{210}$Po activity, the Po removal can be effective. 

In a final series of tests, some alternative cleaning methods are explored to further investigate the role of oxidizers. 
Given there was no added benefit to adding a cell potential to the standard acidified peroxide solution, we attempted to increase the acidity to see if the cell potential could be made to play a larger role. The standard PNNL solution is adjusted to 1M H$_2$SO$_4$ and a slight increase of H$_2$0$_2$ to 6\%. Two separate cleanings were made with this solution while one has an applied cell potential. The results in Fig. \ref{fig:2}(b) show that both cleanings were effective at removing Cu ($\sim 8\,\mu$m of depth) while the very little Po removal occurred, though the sample with the cell potential present hints at slightly more removal of Po. The chemical conditions of this solution provide a competition between the Po and Cu with an advantage to Cu with its lower oxidation potential. The test further confirms that, for this type of solution, there is no advantage to be gained by applying an oxidizing potential when a strong oxidizing agent is present.

To seek some benefit of an oxidizing potential, a test is performed without H$_2$0$_2$ but with a 6M  H$_2$SO$_4$ solution for an even lower pH solution. Again, two samples are cleaned in separate solutions while one sample has an applied cell potential. From the results in Fig. \ref{fig:2}(b), the acidic solution alone was not effective at removing either Cu or Po, while the applied cell potential was sufficient to effectively remove both. This observation suggests that a cell potential alone can oxidize Po in the presence of Cu in an acidic environment, which is consistent with the previous finding of the effectiveness of electropolishing.

\section{OUTLOOK}
There is a rich history of studying Rn progeny deposition and plate-out on surfaces to help guide the design of low background rare event searches. Likewise, successful studies demonstrate methods to remove Rn progeny from metal surfaces, including the problematic $^{210}$Po. Next-generation experiments will have even more stringent  demands for the cleaning and removal of Rn progeny surface contamination. Given the need, it is desirable to use methods that provide both efficient progeny removal and ease of process implementation. 

Several factors are found that determine if the problematic $^{210}$Po progeny will stay in solution during cleaning to be fully removed from a contaminated surface. Oxidizing the Po atoms should keep them in solution and prevent redeposition. As shown here, oxidation of Po can be achieved by an oxidizing agent or an applied potential in the right environment. So the ultimate method to achieve full progeny removal will be dependent on the ideal chemical conditions for the material being cleaned and the chemistry required to achieve and maintain oxidation of the progeny.  In the case of Cu samples, the controlled PNNL acidified peroxide solution is capable of fully removing Po atoms rather than more aggressive methods. However, the cleaning must be performed in a way to give sufficient oxidizing agent exposure commensurate wth the initial $^{210}$Po surface activity. Increased exposure can be achieved through agitation of the samples, greater solution volume in contact with the samples, greater concentration of H$_2$0$_2$, or a longer duration time in solution.

Further studies are needed to fully explore the process variables tested here. The effects of Po solubility, as noted by previous studies \cite{hop07}, will aid in understanding the correlation  between $^{210}$Po activity and Po removal. The role of the kinetics should be explored to understand why a greater exposure is required to oxidize the Po atoms and prevent redeposition than that needed for the substrate atoms. In studies to determine acceptable cleaning solutions, it is desirable to seek out  chemical conditions and processes that favor keeping the progeny in solution.


\section{ACKNOWLEDGMENTS}
We thank our {\sc Majorana}  collaborators for fruitful discussions. 
This material is based upon work supported by the U.S. Department of Energy, Office of Science, Office of Nuclear Physics under Award Number DE-SC0012612. This material is based upon cooperation with S.R. Elliott and Los Alamos National Laboratory.

\bibliographystyle{aipnum-cp}%
\bibliography{/Users/guiseppe/work/latex/bib/myradon}

\end{document}